\documentclass[draft]{agujournal2019}
\usepackage{url} 
\usepackage{lineno}
\usepackage[inline]{trackchanges} 
\usepackage{soul}
\usepackage{float}
\usepackage{caption}
\usepackage{subcaption}

\draftfalse

\journalname{JGR Space Physics}

\begin{document}

\title{ Properties of Mars’ Dayside Low-Altitude Induced Magnetic Field and Comparisons with Venus}

\authors{Susanne Byrd\affil{1}, Zachary Girazian\affil{1}, Suranga Ruhunusiri\affil{2}}

\affiliation{1}{Department of Physics and Astronomy, University of Iowa, Iowa City, IA}
\affiliation{2}{Laboratory for Atmospheric and Space Physics, University of Colorado, Boulder, Boulder, CO, USA}

\correspondingauthor{Susanne Byrd}{susbyrd@uiowa.edu}


\begin{abstract}
Our research objective is to characterize Mars’ low-altitude ($<$ 250 km) induced magnetic fields using data from NASA's MAVEN (Mars Atmosphere and Volatile EvolutioN) Mission. We aim to assess how the induced magnetic fields behave under different solar zenith angles and solar wind conditions, and additionally, understand how planet-specific properties (such as Mars crustal magnetism) alter the formation and structure of the magnetic fields. We then use data from the Pioneer Venus Orbiter to compare induced magnetic fields at Venus with those at Mars. 
At Venus, the vertical structure of the magnetic field tends to exist in one of two states (magnetized or unmagnetized) but we find the induced fields at Mars are more complicated, and we are unable to use this simple classification scheme. 
We also find the low-altitude induced field strength in the ionospheres of both Venus and Mars vary with as cosine of the angle between solar wind velocity and the magnetic pileup boundary.
The low-altitude field strength at Venus tends to be higher than Mars. However, Venus field strengths are lower than theoretical predictions assuming pressure balance and negligible thermal pressure. For Mars, low-altitude field strengths are higher than expected given these assumptions.
Induced field strengths exhibit a trend with solar wind dynamic pressure that is consistent with pressure balance expectations at both planets, however there is significant uncertainty in the Venus fit due to lack of upstream solar wind data.
Our results highlight major differences between the induced magnetic fields at Venus and Mars, suggesting planet-specific properties such as size and the presence of crustal magnetism affect the induced ionospheric magnetic fields at non magnetized planets.

\end{abstract}

\section{Introduction}

A magnetosphere is a region of space around a planetary body where charged particles are deflected by the body’s magnetic field. For bodies with active intrinsic field generation, this region is usually large (e.g., $\sim$10 planetary radii on the sun-facing side for Earth). For bodies without active intrinsic field generation, this region is smaller ($<$ 1 planetary radii) and an induced magnetosphere forms as the magnetic field of the solar wind interacts with the electrically conductive ionosphere \cite{bertucci2011_induced}. As it encounters the obstacle, an electric current is induced in the ionosphere \cite{daniell1977distribution,ramstad_global_2020}. The induced magnetosphere that results from this interaction tends to balance the dynamic pressure from the oncoming solar wind
\cite{luhmann_induced_2004}. Typically, the induced field is much weaker in magnitude than an intrinsic global field.

Induced magnetospheres occur on bodies that have an atmosphere but lack a dynamo-generated global magnetic field. Both Venus and Mars fall into this category as unmagentized planets containing atmospheres, and thus both planets have induced magnetospheres. In this paper, we focus on understanding and comparing the structure and variability of the low-altitude ($<250$ km) induced magnetic fields of Venus and Mars.

The Pioneer Venus Orbiter (PVO) is the only spacecraft to have consistently measured the low-altitude induced magnetic fields at Venus. Analyses of the PVO data confirmed that Venus has a magnetic field induced solely by interaction with the solar wind~\cite{luhmann_intrinsic_1983}. It was also found most vertical profiles of the induced magnetic field strength can, based on their features, be classified as either magnetized or unmagentized \cite{luhmann_observations_1980,luhmann_magnetic_1991}. When the solar wind pressure exceeds the peak ionospheric thermal pressure, the planet's induced field is classified as being in a magnetized state. In the magnetized state, the vertical profile of contains a local minimum near 200 km and a local maximum near 170 km with peak field strengths up to 150 nT. Alternatively, when the solar wind pressure is less than the peak ionospheric thermal pressure, the ionosphere is in an unmagnetized state~\cite{luhmann_magnetic_1991}. In the unmagnetized scenario, the ionosphere excludes most of the external field and the low-altitude induced field strength is weaker ($<30$ nT). These profiles lack distinct maxima or minima at low altitudes, but instead contain many small-scale ($\sim$10 km) spikes characterized as flux ropes \cite{elphic_observations_1980,elphic_1983_growth,elphic_1983_obs_models}.

Some basic characteristics of the low-altitude field strengths in the magnetized ionosphere of Venus have also been reported. The low-altitude field strength was found to decrease with increasing solar zenith angle (SZA) \cite{luhmann_magnetic_1991}. Additionally, the field strength was found to increase with increasing solar wind dynamic pressure \cite{luhmann_observations_1980,kar_mahajan_1987}. Both of these variations are expected if pressure balance is satisfied across the near-space environment and the induced field is required to balance the dynamic pressure of the oncoming solar wind \cite{luhmann_magnetic_1991,sanchez_2020_matter_pressure}. 



Mars Global Surveyor (MGS) provided the first comprehensive magnetic field measurements at Mars, discovering the total field is a combination of induced fields and crustal fields that are scattered across the planet \cite{connerney_global_2001,brain_martian_2003}. However, MGS was unable to measure the induced magnetic field at low altitudes (below 200 km). Arriving at Mars in 2014, the Mars Atmosphere and Volatile EvolutioN (MAVEN) mission became the first spacecraft to routinely measure low-altitude magnetic fields. Recently, \citeA{fang_2023_external_field} analyzed low-altitude magnetic field data from MAVEN. They found the induced field strength on the dayside is usually around 15-50 nT, decreases with increasing SZA, and increases with increasing solar wind dynamic pressure. They also found that the induced field strength is typically stronger than the crustal field strength.  
Several studies have also examined the ubiquitous small scale features (10-50 km) present in the observed magnetic field profiles, including waves, slabs, flux ropes, and flux tubes \cite{hamil_2022_mars_small_scale_structures,bowers_maven_2021,cravens_2023_fourier}. These structures are thought to form through a variety of processes such as variations in the upstream solar wind, global plasma dynamics, and plasma instabilities.


Models have been used in attempt to reproduce the observed low-altitude magnetic field profiles at both planets. For Venus, models from the PVO era \cite{cloutier_1984,cravens_evolution_1984,luhmann_1984,shinagawa_one-dimensional_1988,luhmann_magnetic_1991} and more recent iterations \cite{ma_formation_2020,ma_2023_venus}, have been quite successful at reproducing the observed large-scale magnetic fields present in the magnetized ionosphere. They are able to reproduce several features of the magnetized profiles, including the local minimum near 200 km and the local maximum near 170 km \cite{cravens_evolution_1984,phillips_1984,cloutier_1984,shinagawa_one-dimensional_1988}. Similar models for the induced field at Mars have been somewhat successful at reproducing the observed profiles \cite{shinagawa_one-dimensional_1989,ma_variations_2017,fang_morphology_2018,huang2023variability}. 
Generally, however, the observed induced field profiles at Mars have more small-scale features that models are unable to fully reproduce.

In these models, the evolution of induced magnetic field, $B$, is described by the magnetic diffusion equation:

\begin{equation}
\label{eq:diffuse}
 \frac{\partial B}{\partial t} = \nabla \times (u \times B) - \nabla \times (\eta_{m}\nabla \times B)
\end{equation}

\noindent{where} $t$ is time, $u$ is the plasma flow speed, and $\eta_{m}$ is the magnetic diffusivity. The first term in Equation~\ref{eq:diffuse} represents the convection of magnetic flux with plasma flow and the second term represents the diffusion or dissipation of the magnetic field as the electrical currents associated with the field are dampened through collisions \cite{luhmann_magnetic_1991}. At the top of the ionosphere, which is sometimes called the magnetic pileup boundary or ionopause \cite{espley_2018_terminology}, densities are so low that the first term dominates and the draped solar wind field is convected into the ionosphere by downward flowing plasma. As the field is convected downward, the diffusion term eventually takes over as increased ion-neutral collisions between the induced current and neutral molecules causes the field to dissipate. The vertical structure of the induced magnetic field profile is, to first order, controlled by these processes. An interesting aspect of this picture is that variations in upstream solar wind conditions, such as a change in dynamic pressure, will not have an immediate affect in the low-altitude induced field because the effects take time to propagate through the sheath and into the ionosphere \cite{luhmann_1984,ma_formation_2020,hamil_2022_mars_small_scale_structures,cravens_2023_fourier}.




In this work, we aim to compare the structure and variability of the low-altitude induced magnetic fields at Venus and Mars. Our focus is on large scale structures as opposed to the small-scale structures like flux ropes. Our primary goals are to (1) investigate if the vertical structure of the induced magnetic field profiles at Mars can, like at Venus, be classified as magnetized or unmagnetized; (2) compare the induced field strengths at each planet, and (3) investigate if the low-altitude induced magnetic field strengths vary with SZA and solar wind dynamic pressure in the similar ways.

\section{Data and Method} 

\subsection{Pioneer Venus Orbiter}
PVO collected the bulk of its low-altitude in situ measurements between December 1978 and July 1980 when its orbital period was 24 hours, and periapsis altitude was maintained near 160 km for nearly 700 orbits \cite{brace_structure_1991}. During this time, the periapsis segment evolved to cover a wide range of local times and solar zenith angles, but stayed near mid-latitudes throughout (10$^{\circ}$S-40$^{\circ}$N).

For the magnetic field data, we use the low resolution magnetometer (OMAG) observations that have 12-second time resolution \cite{russell_1980_OMAG}. The files contain the magnetic field vector, spacecraft altitude, solar zenith angle, and latitude of each observation. Each orbit is split into an inbound and outbound segments and trimmed to include only observations at altitudes below 500 km. For upstream solar wind data, we use the upstream solar wind conditions from PVO that were obtained when PVO was located outside the bow shock. The IMF vector was measured by OMAG and the solar wind density and velocity were measured by the Plasma Analyzer instrument \cite{intriligator_pioneer_1980}. The data are provided as hourly averages but we assign a single value to each half-orbit (inbound and outbound) by taking the solar wind measurement closest in time to the low-altitude magnetic field (periapsis) measurement. Periapsis segments without a solar wind measurement within 10 hours were not included. The data set includes solar wind proton densities and velocities from which we calculate the dynamic pressure using $P_{dyn} = \rho V^2$ where $P_{dyn}$ is the solar wind dynamic pressure, $\rho$ is the solar wind mass density, and $V$ is the bulk flow velocity.


\subsection{MAVEN}
 
MAVEN low-altitude in situ measurements cover October 2014 through August 2020 when its orbital period was 4.5 hours and the periapsis altitude varied between $\sim$150-200 km. A few observations come from times when MAVEN was lowered down to $\sim$125 km for week-long “deep dip” campaigns \cite{jakosky_mars_2015}. The location of the periapsis segment slowly evolves throughout the mission, covering a wide range of latitudes, local times, and SZAs. For magnetic field data, we use the Key Parameter (KP) data products. The KP data are a bundle that includes measurements from every MAVEN in situ instrument along with spacecraft ephemeris information, all on a uniformly sampled 4-second time grid. The KP files are compiled from the instrument’s fully calibrated Level 2 data products. As with the Venus data, each orbit is split into an inbound and outbound segment and trimmed to include only observations from below 500 km.

For solar wind data, we use solar wind observations from MAVEN's Magnetometer (MAG) and Solar Wind Ion Analyzer (SWIA) that were obtained while MAVEN is outside the bowshock. The data are derived using the method described in \cite{halekas_structure_2017}. This provides averages of solar wind properties for each orbit. The SWIA observations contain the solar wind proton density and velocity which are used to calculate the solar wind dynamic pressure. Periapsis segments without a solar wind measurement within 10 hours were not included. However, many of the orbits do not have upstream solar wind measurements within 10 hours of the periapsis observations. To fill these gaps, we use the \citeA{ruhunusiri_artificial_2018} solar wind proxy model, which predicts upstream solar wind conditions based on MAVEN observations in the sheath. Of the 2625 MAVEN profiles on the day side, 1151 of them were assigned upstream conditions based on SWIA observations, 1069 were assigned from the proxy model, and 405 could not be assigned because of lack of nearby observations.


\subsection{Caveats}

\begin{figure}
    \centering
    \includegraphics[width=1\linewidth]{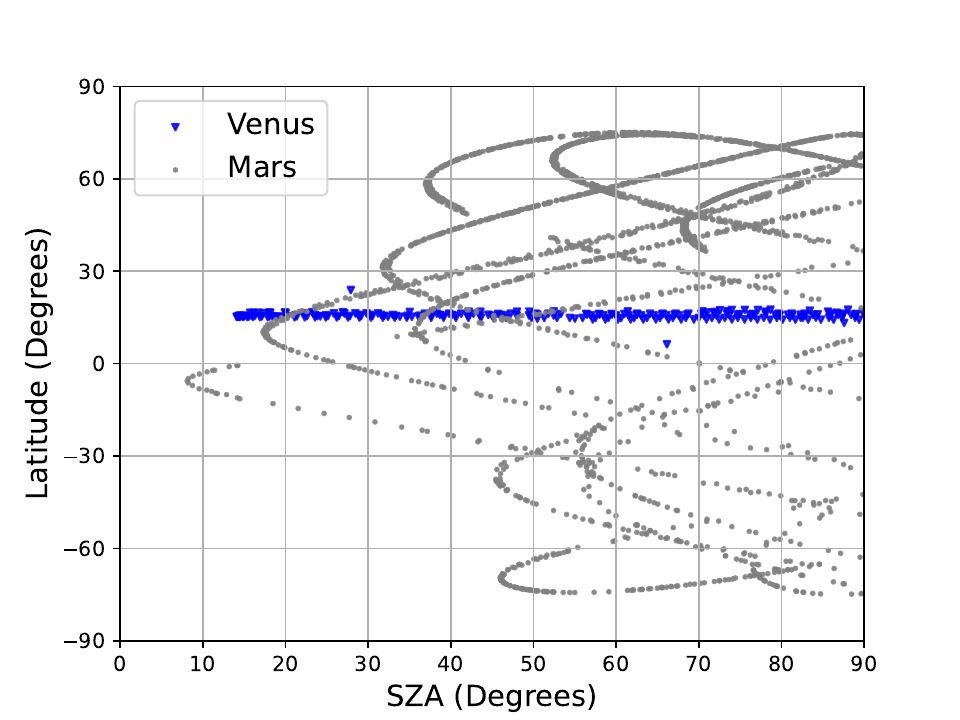}
    \caption{The latitudes and solar zenith angles at periapsis for the orbits used in this study, demonstrating the observational coverage of PVO (blue triangles) and MAVEN (gray circles) are quite different. }
    \label{fig:spacecraft_sza}
\end{figure}

When comparing results derived from the MAVEN and PVO observations, several differences between the two data sets must be considered. First, the PVO data were obtained during the maximum of Solar Cycle 21 \cite{brace_structure_1991}, while the MAVEN data were obtained during the declining phase of the rather weak Solar Cycle 24 \cite{lee_maven_2017}. Second, the PVO periapsis is confined to mid-latitudes, while the MAVEN periapsis precesses between 75$^{\circ}$ N and 75$^{\circ}$ S (See Figure~\ref{fig:spacecraft_sza}). Additionally, because the spacecraft moves both vertically and horizontally during each periapsis pass, the structures we observe in the magnetic field profiles may not strictly be in the vertical direction.

Differences between the planets themselves must also be kept in mind when comparing results from the two planets. For example, Mars has rather pronounced seasons due to its eccentric orbit and 25$^{\circ}$ axial tilt, while Venus has no seasons because of its circular orbit and $<$ 3$^{\circ}$ axial tilt. Other differences to consider include, but are not limited to, planetary rotation rates, the Martian dust season, atmospheric scale heights, and planetary size.

\begin{figure}
\centering{}
\includegraphics[width=1\linewidth]{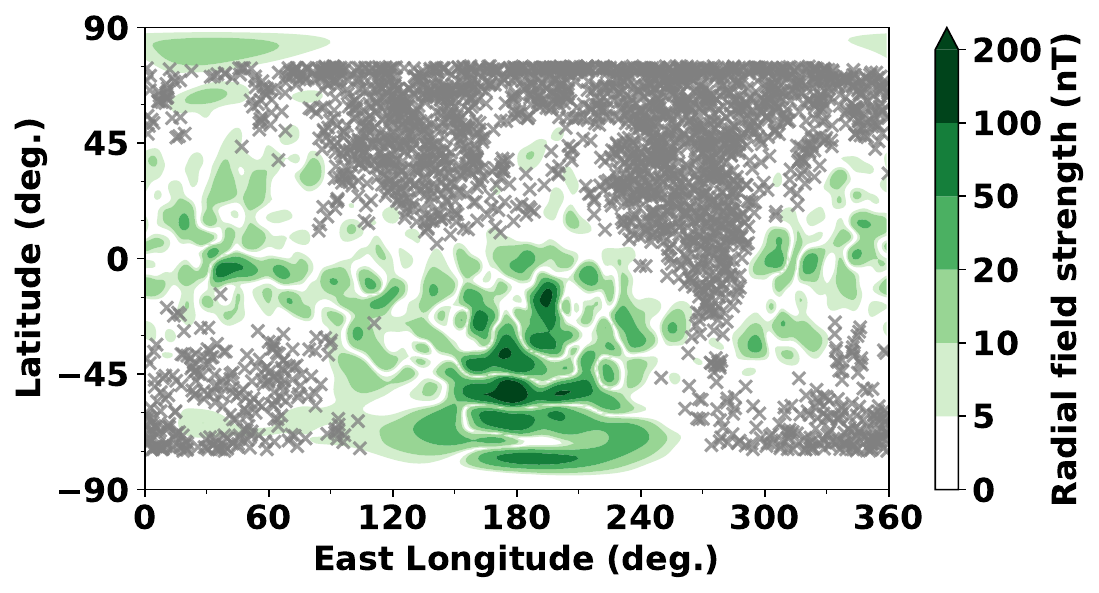}  
\caption{Locations of the MAVEN observations used in our analysis. Grey X's show locations of low-altitude maximum magnetic field strengths ($B_{max}$) per inbound and outbound profile. Green contours are a map of the Martian radial crustal field strength at 400 km \protect\cite{morschhauser_spherical_2014}. These locations are chosen because they are far from strong crustal fields, making them more comparable to Venus.}
\label{fig:crustal}
\end{figure}

\subsection{Method}
\label{sec:method}

The vertical structure of the induced magnetic field profiles at Venus can generally be categorized into one of two states: "magnetized" or "unmagnetized". We categorize the Venus profiles in this manner based on the descriptions in \citeA{luhmann_magnetic_1991}. If the global minimum of a magnetic field profile is below 300 km, and the profile has low-altitude ($<250$ km) maximum field strength greater than 30 nT, we classify it as magnetized. Or, if a profile has a low-altitude magnetic field strength greater than 50 nT, we classify it as magnetized. If a profile has a low-altitude maximum field strength less than 30 nT, it is classified as unmagnetized. A small number of profiles remain categorized because they they do not match either criteria.

\begin{figure} 
    \centering
    \includegraphics[width=1\linewidth]{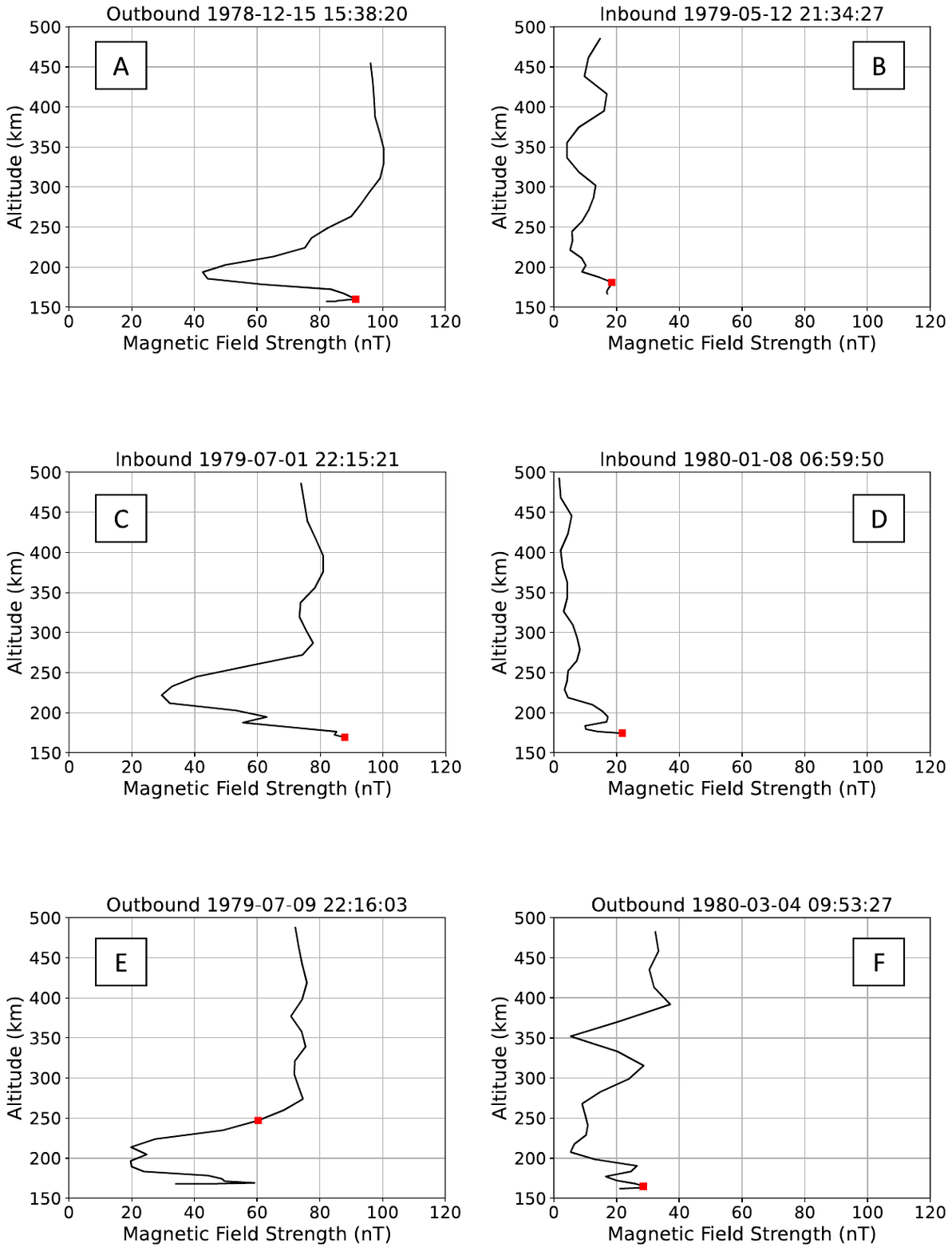}
\caption{Examples of magnetic field profiles from Venus with magnetized states in the left column and unmagnetized states in the right column. The squares mark the derived $B_{max}$ for each profile. }
    \label{fig:venus_profiles}
\end{figure}

Figure~\ref{fig:venus_profiles} shows six examples of Venus magnetic field profiles observed by PVO. The examples in the left column show magnetized profiles. They contain a peak below 200 km, a wide minimum near 200-250 km, and a nearly-constant topside. These features are all absent in unmagnetized profiles, examples of which are shown in the right column of Figure~\ref{fig:venus_profiles}. Instead, the unmagnetized profiles are nearly featureless, lacking distinct maxima or minima, and have weaker field strengths at nearly all altitudes.

\begin{figure} 
    \centering
    \includegraphics[width=1\linewidth]{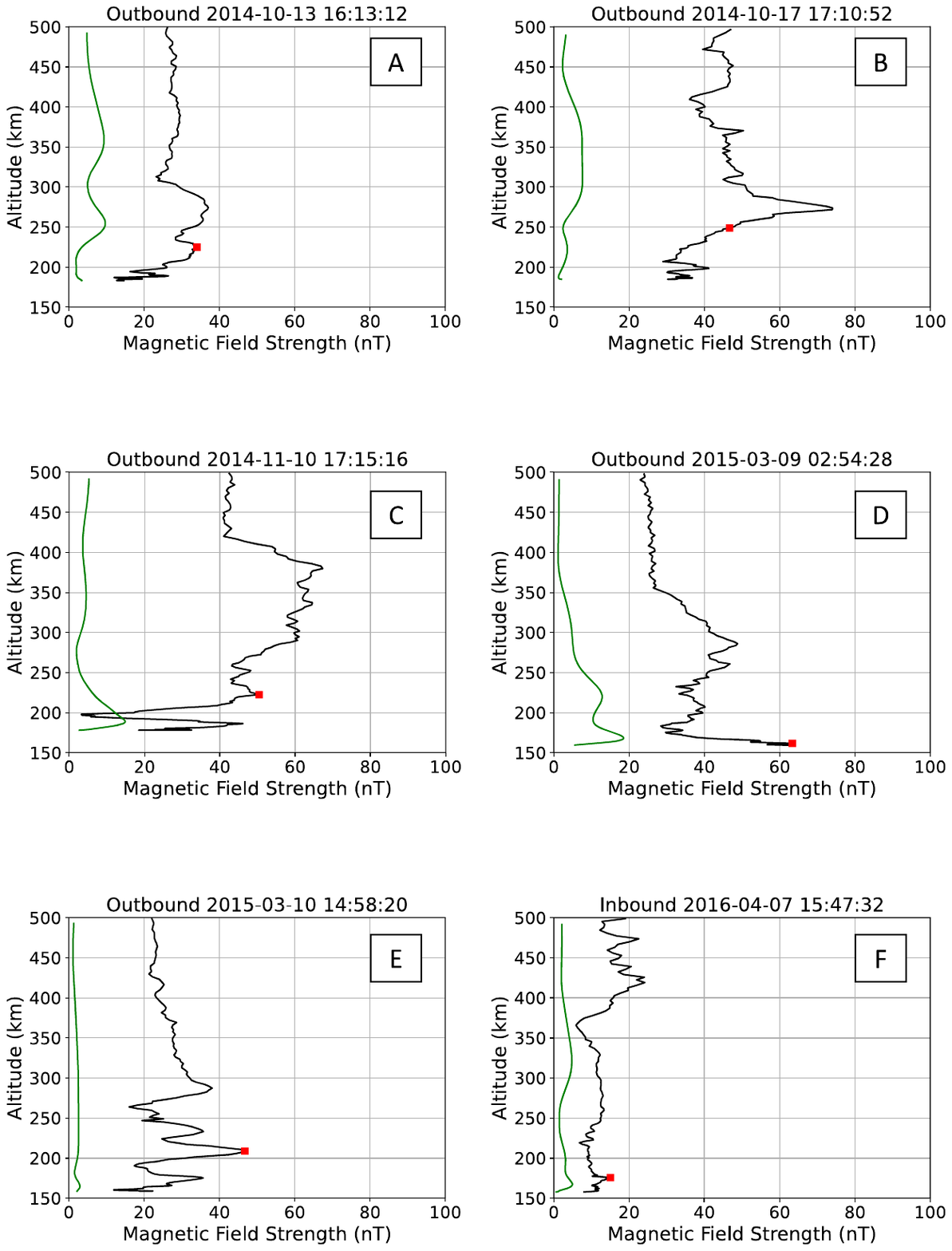}
        \caption{Examples of magnetic field profiles from Mars. The black lines are the measured magnetic field from MAVEN and the green lines are the crustal magnetic field strength from the \protect\citeA{morschhauser_spherical_2014} crustal field model. The red squares mark the derived $B_{max}$ for each profile.  }
\label{fig:mars_profiles}
\end{figure}

Figure~\ref{fig:mars_profiles} shows six examples of magnetic field profiles observed by MAVEN. These profiles are much more complicated than profiles from Venus and they lack recognizable features that would allow them to be sorted easily into two categories. This issue will be explored further in Section~\ref{sec:classification}.


Since we are interested in comparing the strength and variability of the low-altitude induced fields at Venus and Mars, we focus our analysis on the maximum magnetic field strength below 250 km altitude ($B_{max}$). To derive $B_{max}$ for both inbound and outbound segments of an orbit, we locate the maximum field strength below 250 km. 
The $B_{max}$ derived from each profile Venus profile is marked in Figure~\ref{fig:venus_profiles}. For magnetized profiles, $B_{max}$ is the field strength at the location of the low-altitude peak near 170 km. For unmagnetized profiles, $B_{max}$ is the maximum field strength below 250 km, but is not at the location of a specific feature.

We use the same method to extract $B_{max}$ from the Mars profiles. Figure~\ref{fig:mars_profiles} shows the derived $B_{max}$ locations for six example profiles. Since the vertical structure varies much more from profile to profile, the derived $B_{max}$ is not extracting the peak strength from a recurring distinct feature, as in the case of the magnetized Venus profiles.

We note that there are several cases when the location of the derived $B_{max}$ is either at the bottom of the profile (periapsis) or at the top of the profile (near the top altitude cutoff of 250 km). For cases when the derived $B_{max}$ is near the high altitude cutoff, the low-altitude field strength never exceeds the field strength at 250 km. Examples of these cases can be seen in Figure~\ref{fig:venus_profiles}E and Figure~\ref{fig:mars_profiles}B. For these, $B_{max}$ is likely overestimated because our method does not pick any low-altitude feature that has a local maximum below 250 km. For cases when $B_{max}$ is near periapsis, the derived $B_{max}$ likely underestimates the true local maximum in the low-altitude magnetic field profile. The maximum likely occurs below periapsis where there are no observations. Examples of these profiles can be seen in Figure~\ref{fig:venus_profiles}C and Figure~\ref{fig:mars_profiles}D. Nonetheless, we include these cases in our final dataset.

 Another important consideration is the crustal magnetic fields at Mars that are not at Venus. We wish to remove these crustal fields from our analysis to enable a more direct comparison between the two planets, focusing solely on the induced component of the magnetic field. We use the \citeA{morschhauser_spherical_2014} empirical model of crustal magnetic field strength to omit MAVEN observations where the crustal field strength is comparable to the induced field strength. More concretely, we exclude any profiles that have $B_{crust}/B_{max}$ $>$ 0.2 where $B_{crust}$ is the crustal field strength at the location of $B_{max}$. Figure~\ref{fig:crustal} shows the locations of $B_{max}$ that remain after applying this crustal field filtering. It demonstrates the chosen $B_{max}$'s are confined to outside the crustal field regions. 

To focus on the dayside interaction region, in our final dataset we only use profiles that have $B_{max}$ at SZA $< 90^{\circ}$. In total, the final filtered dataset for Venus includes 475 magnetic field profiles of which 188 of which are categorized as magnetized, 260 as unmagentized, and 27 as uncategorized. Of the 188 magnetized profiles, 39 have $B_{max}$ at the bottom of the profile and 25 have $B_{max}$ at the top altitude cutoff of 250 km. Of the 260 unmagentized profiles, 6 have $B_{max}$ at the bottom of the profile and 30 have $B_{max}$ at 250 km. The final filtered dataset for Mars includes 2625 profiles. Of these, 220 have $B_{max}$ at the bottom of the profiles, and 112 have $B_{max}$ at the top altitude cutoff of 250 km.


\section{Analysis}

\subsection{Categorizing the Magnetic Field Profiles}
\label{sec:classification}

As discussed in Section~\ref{sec:method} and shown in Figure~\ref{fig:venus_profiles}, the induced magnetic field profiles at Venus have reoccurring vertical structures that can generally be categorized as either "magnetized" or "unmagnetized" \cite{luhmann_magnetic_1991}. Using our classification criteria, we find $\sim$40\% of the profiles can be classified as magnetized and $\sim$55\% can be classified as unmagnetized. These percentages are loosely consistent with the occurrence rates reported by \citeA{luhmann_observations_1980}, who categorized 30\% as magnetized and 70\% as unmagnetized. However, they used somewhat different classification criteria. In this section, we wish to determine if the magnetic field profiles at Mars can be classified in the same way.

We attempted to categorize the Mars profiles based on visual inspection but quickly found they could not be categorized into Venus-like magnetized and unmagentized vertical structures. This is demonstrated in Figure~\ref{fig:mars_profiles}, which shows six example profiles from Mars. These six profiles lack distinct features that would enable them to be easily categorized. In particular, they lack a consistent minimum near 300 km or a clear single low-altitude peak, features that are present in the Venus magnetized profiles. Instead, the Mars profiles are more varied and have more complex small-scale structures (tens of kilometers). These structures are present even when the crustal field component is much lower than induced field strength.

We conclude the vertical structure of the induced magnetic fields at Mars are too complex to be classified into simple magnetized and unmagnetized states. Further, it is generally accepted that Mars is magnetized more often than Venus, meaning the ionospheric thermal pressure is usually insufficient to exclude the external field from penetrating into the ionosphere.\cite{zhang_comparisons_1992,chu_2021_ionopause,holmberg_maven_2019,sanchez_2020_matter_pressure}. We find the profiles resembling the magnetized ionosphere of Venus, such as the profile shown in Figure~\ref{fig:mars_profiles}C, do exist but are quite rare. Some of the profiles, such as Figure~\ref{fig:mars_profiles}F, also have a similar appearance to the unmagnetized profiles at Venus (Figure~\ref{fig:venus_profiles} B, D, and F). Namely, low magnetic field strength at all altitudes and a lack of a distinct large scale peak. 


Although the induced magnetic field profiles at Mars are unable to be categorized into simple states like the case at Venus, there are some reoccurring features that are worth reporting. First, the profiles often have a distinct prominent peak between 250-300 km. Examples of such profiles can be seen in Figure~\ref{fig:mars_profiles}A, B, and D. 


Some profiles have many peaks, such as the profile in Figure~\ref{fig:mars_profiles}E. These multiple peaks at low altitudes may be due to a combination of both horizontal and vertical variations in the magnetic field strength (the spacecraft trajectory is not strictly vertical). Generally, magnetized profiles at Venus lack such a complicated structure. 

Neither the complicated multi-peaked structures, nor a peak between 250-300 km were predicted by the earliest MHD models~\cite{shinagawa_one-dimensional_1989}. More recent MHD studies that include the time-dependent effects of changing solar wind conditions do predict more complicated structures including a peak between 250-350 km  \cite{ma_martian_2014,ma_variations_2017}. The peaks are caused by vertical gradients in the downward plasma flow speed as a result of solar wind pressure variations on hour long timescales. Solar wind variations may be responsible for some of these small scale features \cite{cravens_2023_fourier}, but other processes such as plasma instabilities and global plasma dynamics likely play a role \cite{hamil_2022_mars_small_scale_structures}.


\subsection{Solar Zenith Angle Variations}
\label{sec:sza_variations}

To first order, we expect pressure balance to be maintained through the space environment. Given this assumption, the maximum induced magnetic field strength should depend on the solar wind dynamic pressure and the ionospheric thermal pressure. Higher solar wind dynamic pressures should lead to stronger maximum field strengths,  while higher ionospheric thermal pressures should be able to exclude the solar wind field more efficiently and result in weaker induced magnetic fields.

If pressure balance is satisfied throughout the low-altitude ionosphere then

\begin{equation}
\label{eq:pressure_balance}
    P_{sw}cos^{2}{(\chi)} = P_{th} + P_B
\end{equation}

\noindent{where} $P_{sw}cos^{2}{\chi}$ is the normal component of the solar wind dynamic pressure \cite{crider_proxy_2003}, $\chi$ is SZA, $P_{th}$ is the ionospheric thermal pressure, and $P_B = B^2_{ind}/{2\mu_o}$ is the induced magnetic field pressure. If we assume the $P_B >> P_{th}$, which is often the case \cite{holmberg_maven_2019,sanchez_2020_matter_pressure}, then we expect the induced field strength to approximately follow 

\begin{equation}
\label{eq:cosine}
    B_{ind} = (2\mu_oP_{sw})^{1/2}cos{(\chi)}
\end{equation}

\noindent{where} $\mu_o$ is the vacuum permeability. This equation predicts that, when magnetized ($P_B > P_{th}$), the induced magnetic field strength should be approximately proportional to cosine of the SZA. We examine this prediction for both Venus and Mars.

\begin{figure} 
    \centering
    \includegraphics[width=1\linewidth]{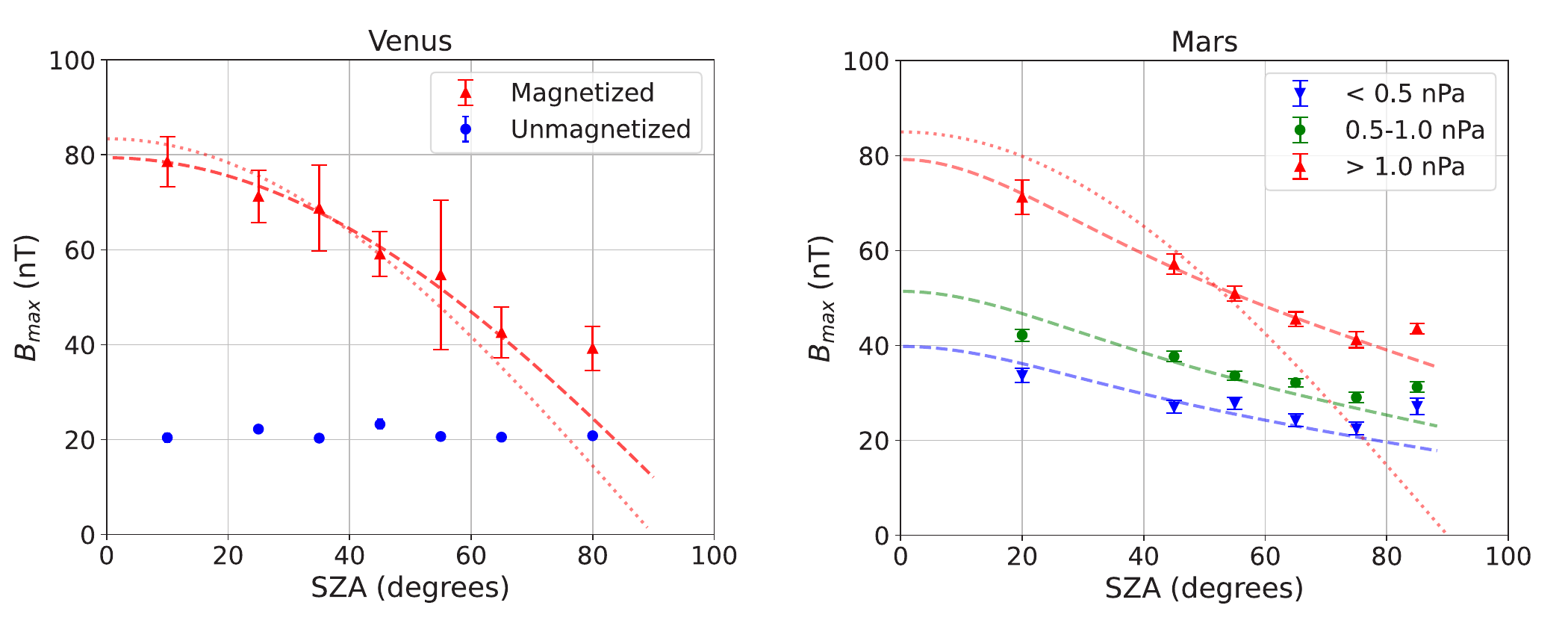}
    \caption{The solar zenith angle (SZA) variation of $B_{max}$ for Venus (left) and Mars (right). The dotted lines are fits to $B_{max} = B_{0}cos(\chi)$ (see Equation~\ref{eq:cosine}) using SZA for $\chi$. The dashed lines are fits using $\theta$ (the angle between the solar wind flow and the magnetic pileup boundary) in place of $\chi$ \protect\cite{vignes_2000_mars_mpb_shape,xu_2021}. The Venus data are sorted into magnetized and unmagnetized cases while the Mars data are sorted into solar wind dynamic pressure bins. Error bars show the standard error in each bin. The solar zenith angle bins for Venus have edges at 0$^{\circ}$, 20$^{\circ}$, 30$^{\circ}$, 40$^{\circ}$, 50$^{\circ}$, 60$^{\circ}$, 70$^{\circ}$, and 90$^{\circ}$. The SZA bins for Mars have edges at 0$^{\circ}$, 40$^{\circ}$, 50$^{\circ}$, 60$^{\circ}$, 70$^{\circ}$, 80$^{\circ}$, and 90$^{\circ}$.} 
    \label{fig:sza_trends_combined}
\end{figure}

Figure~\ref{fig:sza_trends_combined} shows $B_{max}$ plotted against SZA for both planets. The Venus observations are separated into their magnetized and unmagnetized classifications, but not separated based on solar wind dynamic pressure because of data scarcity. The Mars observations are sorted into groups based on solar wind dynamic pressure since the induced field will increase with increasing solar wind pressure (see Section~\ref{sec:psw}). The data are binned along the SZA axis using uneven bin sizes so that each bin has a comparable number of data points.


At Venus the maximum field strengths of the unmagnetized profiles do not have a SZA dependence and are consistently around 20 nT, which is consistent with previous studies \cite{elphic_venus_1984}. This is expected since when the ionosphere is unmagnetized, the external magnetic field is excluded from the ionosphere by the high ionospheric thermal pressure, and thus the induced field strength is not expected to be driven by the solar wind pressure. In these cases $P_B << P_{th}$ so Eq.~\ref{eq:cosine} is not applicable and we do not expect any cosine dependence. 
In contrast, the maximum field strengths of the magnetized profiles decrease with increasing SZA up to a SZA of $\sim$65$^{\circ}$. The bin-averaged maximum field strength decreases from $\sim$80 nT near the subsolar point to $\sim$40 nT near 65$^{\circ}$ SZA. 

Similarly, the maximum field strengths at Mars decrease from $\sim$70 nT (35 nT) at the subsolar point to $\sim$40 nT (25 nT) around 75$^{\circ}$ SZA during high (low) solar wind dynamic pressure. There is also clear separation between the dynamic pressure bins, with higher maximum field strengths occurring during higher solar wind dynamic pressures, as predicted by Equation~\ref{eq:cosine}. The solar wind dynamic pressure trends at both planets will be further explored in Section~\ref{sec:psw}.

The dotted lines are fits to the data using SZA for $\chi$ in Equation~\ref{eq:cosine}). Specifically, we fit the binned averages to $B_{max} = B_{0}cos(\chi)$ where $B_0$ is a fit parameter. In both cases, the data are poorly fit at high SZAs ($>$ 60$^\circ$) because $B_{max}$ does not decrease with SZA as rapidly as the cosine curve predicts. We suggest this deviation from the prediction is a consequence of SZA being only an approximation of the angle between the solar wind bulk flow velocity and the obstacle. We refit the data after replacing $\chi$ with $\theta$, where $\theta$ is the angle between the solar wind flow and the MPB. For Venus we use the MPB shape found by~\citeA{xu_2021} and for Mars we use the MPB shape given in \citeA{vignes_2000_mars_mpb_shape}.

We fit the binned averages to $B_{max} = B_{0}cos(\theta)$ where $B_0$ is a fit parameter. For Venus the fitted $B_0$ is 79.5$\pm$0.9. For Mars, the fitted $B_0$ is 79.2$\pm$0.4, 51.7$\pm$2.0, and 40.1$\pm$1.4 for the high, medium, and low solar wind bins, respectively. We find this model to be an improved fit to the data, as it follows closely with the data even at higher SZA. It is especially apparent at Mars that the data follows the shape of the MPB rather than a cosine of SZA. However, at both planets $B_{max}$ deviates from the model shape of the MPB near the terminator.

From these two plots we conclude low-altitude maximum field strength follows the predicted cosine trend (from Equation~\ref{eq:cosine}) up to at least 60$^{\circ}$ when in a magnetized state. But only if the angle between the solar wind flow and the MPB is used ($\theta$). From this we can conclude that the pressure balance relation (Equation~\ref{eq:pressure_balance}) is satisfied at lower altitudes than the MPB. However, at SZA near the terminator $B_{max}$ is higher than we expect at both planets. We currently do not have an explanation for this unexpected behavior.


\subsection{Peak Magnetic Field Distributions}
\label{sec:histograms}

\begin{figure} 
    \centering
    \includegraphics[width=1\linewidth]{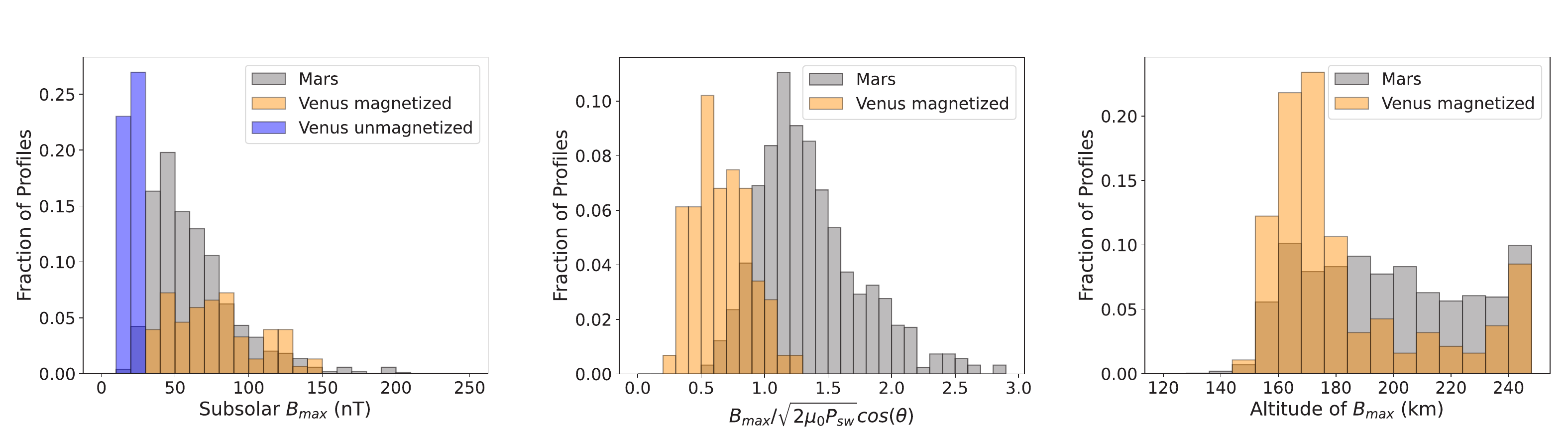}
    \caption{Left: Histograms of $B_{max}$ (maximum low-altitude field strength). Middle: Ratios of the observed $B_{max}$ and the $B_{max}$ predicted by Equation~\ref{eq:cosine}. Right: Histograms of the altitude of $B_{max}$. } 
    \label{fig:hist}
\end{figure}

In Figure~\ref{fig:hist}a, we plot histograms of $B_{max}$ with the Venus data separated into magnetized and unmagnetized categories (Section~\ref{sec:classification}). All the $B_{max}$ values from Mars and all the magnetized $B_{max}$ values from Venus are corrected to the subsolar point by dividing them by $cos{(\mathrm{\theta})}$. Only $B_{max}$ values at SZA $<$ 60$^{\circ}$ are included (Section~\ref{sec:sza_variations}). 

The histogram of the unmagnetized $B_{max}$ values at Venus form a sharp peak around 20 nT, which is the typical magnitude of the induced field when unmagnetized. The histogram of the magnetized $B_{max}$ values at Venus form a much wider distribution, but demonstrates the clear demarcation between magnetized and unmagentized distributions. The magnetized distribution has a mean of 77.8 ($\pm$29.2) nT and a median of 75.7 nT. Comparatively, the distribution of the Mars $B_{max}$ values has a mean of 62.9 ($\pm$29.4) nT and a median of 55.7 nT.

At the subsolar point, Equation~\ref{eq:cosine} predicts $B_{max} =(2\mu_oP_{sw})^{1/2}$ \cite{fang_2023_external_field}. Since $P_{sw}$ scales as the inverse square of the planet-Sun distance, we expect larger $B_{max}$ values at Venus than Mars, which is confirmed in Figure~\ref{fig:hist}. Further, since the solar wind dynamic pressure scales with distance as an inverse square law, Equation~\ref{eq:cosine} predicts $\frac{B_{max,\mathrm{Venus}}}{B_{max,\mathrm{Mars}}} \simeq \frac{1.5 \mathrm{AU}}{0.7 \mathrm{AU}} \simeq 2.1$. However, the observed ratio is only 1.2 if using the means of the distributions  (77.8 nT/62.9 nT), or 1.4 if using the medians (75.7 nT/55.7 nT). If we account for the crustal field component at Mars, that is at most 20\% of $B_{max}$ (in most cases is less than 10\% of $B_{max}$), then the observed ratio is at most 1.75 (1.4/0.8), implying the observed ratio is smaller than the predicted ratio of 2.2.

To explore why the ratio is smaller than predicted, Figure~\ref{fig:hist} b shows a comparison of the observed and predicted $B_{max}$. In particular, the histograms show the ratio between the observed $B_{max}$ and the predicted $B_{max}$ for each profile.  Only observations with SZA $< 60^{\circ}$ are included. The Venus ratios are usually less than one, having a mean of 0.66 ($\pm$0.22) nT. A possible explanation is that Equation~\ref{eq:cosine} relies on two assumptions: (1) pressure balance throughout the space environment and (2) the ionospheric thermal pressure is negligible compared to the magnetic pressure. At Venus, the second assumption is often not satisfied \cite{luhmann_characteristics_1987} leading to a reduction in the induced field strength via Equation~\ref{eq:pressure_balance}.

At Mars, the ratios are usually greater than one, with a mean of 1.38 ($\pm$0.22) nT. If we consider a 20\% crustal field component, the mean would decrease to at least 1.1, still implying the induced field strengths at Mars are somewhat larger than predicted by Equation~\ref{eq:cosine}. For Mars, the second assumption is usually satisfied \cite{holmberg_maven_2019,chu_2021_ionopause,sanchez_2020_matter_pressure} and so it might be expected that Equation~\ref{eq:cosine} is a good predictor of the maximum induced field strength. A possible explanation for why $B_{max}$ is usually larger than the prediction is the presence of small-scale features in the magnetic field profile. These small scale features are much more prevalent at Mars than at Venus. They result in small-scale spikes and bumps that can lead to a higher $B_{max}$ than predicted. These small scale structures arise from a variety of processes, such as waves, rotations, plasma instabilities \cite{hamil_2022_mars_small_scale_structures,cravens_2023_fourier}.



Lastly, Figure~\ref{fig:hist}c shows histograms of the altitude where $B_{max}$ is observed. At Venus, the altitude of the low-altitude maximum induced field, when magnetized, is consistently around 170 km. 
There are rarer cases where the altitude of $B_{max}$ is much higher, including a large number of profiles with the altitude bin in the highest histogram bin near 250 km. These are the edge cases discussed in Section~\ref{sec:histograms} and are profiles similar to Figure~\ref{fig:venus_profiles}E.

The peak altitude at Mars is typically between 160-200 km, but the distribution is much wider than the distribution at Venus. The distribution is likely wider because of seasonal variations in the upper atmosphere of Mars, which causes constant pressure levels to rise and fall over the year owing to the changing insolation from Mars' elliptical orbit.

\subsection{Solar Wind Dynamic Pressure}
\label{sec:psw}

It is expected that with increasing solar wind pressure, the induced field strength will also increase. Assuming pressure balance across the magnetic pileup boundary and into the ionosphere, and negligible ionospheric thermal pressure, the magnetic field should vary as the square root of solar wind dynamic pressure (Equation~\ref{eq:cosine}). Previous work has shown that the magnetic field at the MPB does respond in this manner to solar wind dynamic pressure variations ~\cite{crider_proxy_2003,xu_2021}. Below the MPB, we still expect the induced field to increase with increasing dynamic pressure as magnetic fields induced near the boundary are pulled down with the downward moving plasma. 

\begin{figure} 
    \centering
    \includegraphics[width=1\linewidth]{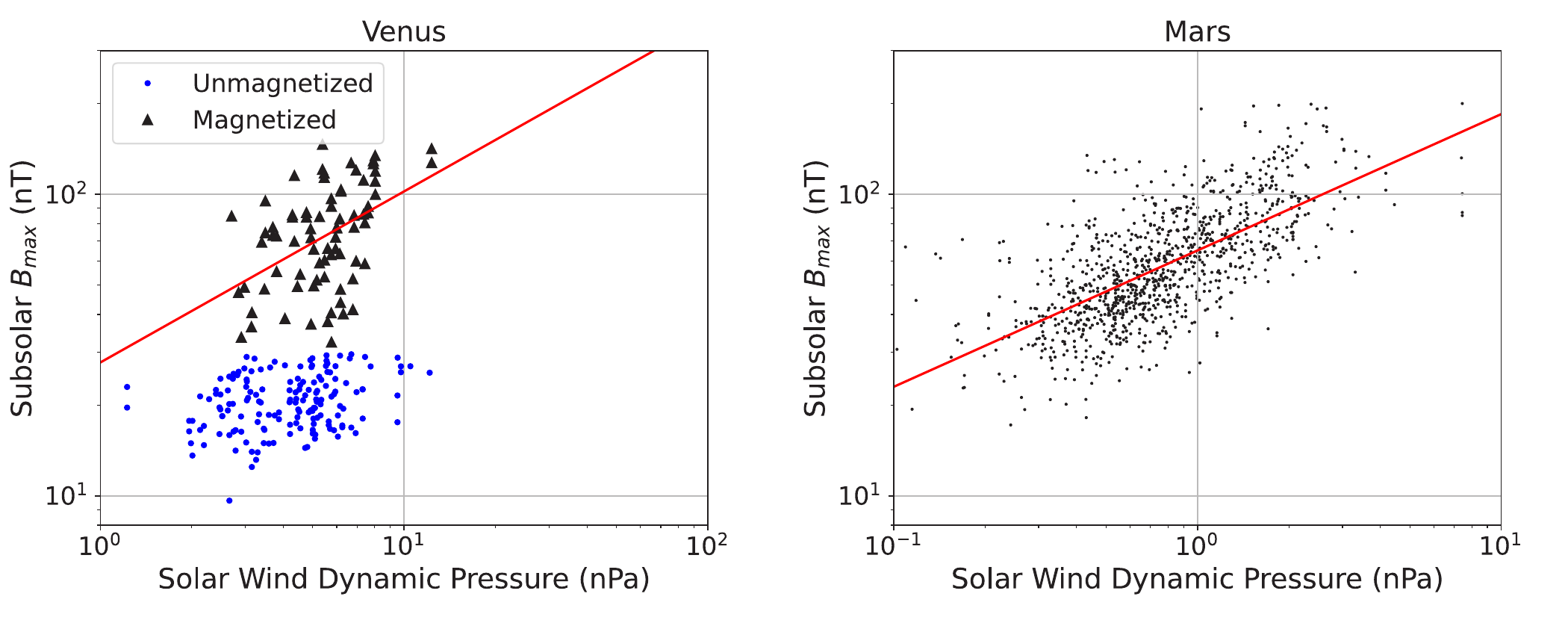}
    \caption{The variation of $B_{max}$ with solar wind dynamic pressure for Venus (left) and Mars (right). For Venus, values derived from magnetized profiles are plotted as black triangles and values derived from unmagnetized profiles are plotted as blue circles. The red lines are power law fits with slopes 0.57$\pm$0.13 for Venus (magnetized) and 0.45$\pm$0.02 for Mars.  }
    \label{fig:sw_plots}
\end{figure}

In figure~\ref{fig:sw_plots} we plot $B_{max}$ versus solar wind dynamic pressure. The magnetized Venus data and the Mars data were corrected to the subsolar point by dividing by $cos(\theta)$ (see Section~\ref{sec:sza_variations}). The plot only contains data with SZA $<60^{\circ}$ since the data deviates from the $cos(\theta)$ trend after that. We fit the following power law to the data: $B_{max} = B_0 P_{sw}^n$ where $B_0$ and $n$ are fitted parameters, $B_{max}$ is low-altitude induced field strength, and $P_{sw}$ is solar wind dynamic pressure.

We find that $B_{max}$ does increase as solar wind pressure increases for both Venus (when magnetized) and Mars with power law exponents $n$ being 0.57$\pm$0.13 and 0.45$\pm$0.02 respectively. These are consistent with each other within error, however the Venus fit has significant uncertainty due to having much fewer data points and a more narrow range of solar wind pressure measurements compared to Mars.


The coefficient $B_0$ of the power law fit for Venus (magnetized) is 27.7$\pm$6.1 whereas the coefficient for Mars is 65.2$\pm$0.7. This implies that for a solar wind pressure of 1 nPa, Mars would have a higher induced field, which is expected since Venus tends to exclude the external magnetic field more effectively than Mars (consistent with figure~\ref{fig:hist}c). However, Venus is more likely to experience higher solar wind dynamic pressure since it is closer to the Sun. As a result, the average $B_{max}$ for Venus (when magnetized) is higher than at Mars (see Section~\ref{sec:histograms}). However, we note that there is significant uncertainty in the Venus fit (R squared values for Venus and Mars are 0.2 and 0.6 respectively). The slope for the unmagnetized Venus data is virtually zero, with a low R squared value of 0.01, as expected since when unmagnetized the solar wind magnetic field is excluded from the low-altitude space environment.




\section{Discussion and Conclusions}


In this study we investigate low-altitude induced magnetism at Mars in comparison to Venus. We show that:

\begin{enumerate}
    \item The magnetic field profiles at Venus and Mars have different altitude structures. Venus profiles typically can be categorized as magnetized or unmagnetized. The magnetized profiles have distinct, identifiable features, namely a peak in magnetic field around 170 km, a wide minimum around 200 km, and a nearly constant topside above 250 km. The unmagnetized profiles do not have large scale features, and have low magnetic field strengths at all altitudes. These findings are consistent with~\citeA{luhmann_magnetic_1991}. When attempting to categorize Mars profiles the same way, we find that they lack these distinct repeatable large-scale features that would allow them to be classified into the same categories as Venus. 

    \item The maximum low-altitude magnetic field strength ($B_{max}$) from unmagnetized profiles at Venus have no dependence on SZA. $B_{max}$ at Mars and Venus (in a magnetized state) exhibit a SZA dependence that is consistent with pressure balance expectations. Namely, $B_{max}$ decreases as $\cos(\theta)$, where $\theta$ is the angle between the solar wind velocity and the MPB normal vector~\cite{vignes_2000_mars_mpb_shape,xu_2021}. However, at both planets, $B_{max}$ is larger near the terminator than this trend predicts.

    \item On average, Venus $B_{max}$ is greater than Mars (with means of 77.8 nT and 62.9 nT, respectively) which is expected since Venus experiences higher solar wind pressure. However, when magnetized, average $B_{max}$ at Venus is lower than predicted from the pressure balance and $P_{th} << P_B$ assumptions by a factor of 0.66. With Mars on the other hand, average $B_{max}$ is higher than predicted by a factor of 1.38. Even after considering maximum possible contributions from crustal magnetic fields, this is still higher than predicted by a factor of 1.1.

     \item When unmagnetized, Venus $B_{max}$ shows no trend with solar wind dynamic pressure ($P_{sw}$). For magnetized profiles on Venus and Mars profiles, there is a trend with $P_{sw}$ (slopes in log-log space 0.57$\pm$0.13 and 0.45$\pm$0.02 respectively). This is consistent within error to the prediction from pressure balance assumptions (Eq.~\ref{eq:cosine}).  


\end{enumerate}



The vertical structure of the magnetic field profiles at Mars is highly varied and inconsistent, in contrast with the Venus profiles which tend to have the same recurring large-scale features. A likely explanation is the ubiquitous presence of small-scale structures in the magnetic field profiles at Mars, which are much less common in the magnetized ionosphere of Venus. This explanation was also put forth by \citeA{huang2023variability} to explain the varied magnetic field profiles at Mars. They attributed the varied structures to time-dependent variations in the up stream solar wind, such as variations in the solar wind dynamic pressure. 
The vertical structure of the Venus altitude profiles can be successfully reproduced by models, even under steady solar wind conditions \cite{luhmann_1984,cravens_evolution_1984,shinagawa_one-dimensional_1988,luhmann_magnetic_1991}. However, models have been less successful for Mars as the vertical structures are much more varied. Additional physical processes need to be considered to be accurate at Mars. ~\citeA{cravens_2023_fourier} presented strong observational evidence of low-altitude magnetic field fluctuations in response to the solar wind. 


We find both similarities and differences between the low-altitude magnetic field strengths. The main similarity arises between Mars and the magnetized ionosphere of Venus, providing further evidence that the ionosphere of Mars is magnetized most of the time \cite{sanchez_2020_matter_pressure}. 
In particular, the low-altitude field strengths have the same SZA dependence, including higher than expected fields near the terminator. We currently do not have an explanation for this deviation. Nonetheless, at both planets the SZA variation of the low-altitude field strengths are modeled best when the angle between solar wind velocity and the normal vector of the MPB is used. 
%
%
This suggests that the solar wind's interaction with this boundary (as opposed to the bowshock) likely controls the induced field strength at low altitudes.

Major differences include higher induced field strengths in the magnetized ionosphere of Venus (average of 77.8 nT vs. 62.9 nT at Mars), which is expected given Venus is exposed to higher solar wind dynamic pressures due to its proximity to the Sun. However, we find that when magnetized, the low-altitude field at Venus is weaker than predicted under the assumptions of pressure balance and $P_{th} << P_B$. This implies the ionospheric thermal pressure at Venus is non-negligible in many cases 
and consequently the induced field does not have to be as large to maintain pressure balance (Eq.~\ref{eq:pressure_balance}).

In contrast, on Mars, low-altitude field strengths are higher than predicted, even after taking into consideration a possible contribution from crustal fields. We currently do not have an explanation for this, but it may be related to internal processes at Mars that are uncommon at Venus, such as small scale features, instabilities, current sheets, etc., and these processes likely originate from planetary property unique to Mars. 


The final difference is the trend between $B_{max}$ and $P_{sw}$. The exponent of our power law fit to these data are 0.57$\pm$0.13 for Venus (magnetized) and 0.45$\pm$0.02 for Mars. These are consistent within error to our pressure balance assumption (see Section~\ref{sec:psw}), meaning pressure balance is satisfied at lower altitudes than the MPB. Although for Venus we acknowledge there is significant error in the fit.

As a whole our results point to same interesting similarities and differences in the induced magnetic fields at Venus and Mars. They point to how planet specific properties such as planetary size, presence of crustal magnetism, or other internal processes can affect induced field structure and variability on unmagnetized planets.

\section{Open Research}
The data used in this work are publicly archived. 


\acknowledgments
This work was supported by NASA grant 80NSSC21K0147 which was funded through the Solar System Workings program. S. Byrd received additional support from the NASA Iowa Space Grant Consortium and the John and Elsie Mae Ferentz Undergraduate Research Fund.



\end{document}